\documentclass{article}
\usepackage{graphicx} % Required for inserting images
\usepackage[T1]{fontenc}
\usepackage[space]{cite}

\title{Preface to Fields, Gravity, Strings and Beyond: In Memory of Stanley Deser}
\author{Marc Henneaux\footnote{Universit\'e Libre de Bruxelles and International Solvay Institutes, ULB-Campus
Plaine CP231, B-1050 Brussels, Belgium \quad {\tt marc.henneaux@ulb.be}} \footnote{Coll\`ege de France, Universit\'e PSL, 11 place Marcelin Berthelot, 75005 Paris, France}, 
Rafael I. Nepomechie\footnote{Department of Physics, PO Box 248046, University of Miami, Coral Gables, FL 33124 USA \quad {\tt nepomechie@miami.edu}}, 
and Domenico Seminara\footnote{Dipartimento di Fisica, Universit\`a di Firenze and INFN Sezione di Firenze, via G. Sansone 1, 50019 Sesto Fiorentino, Italy \quad {\tt domenico.seminara@unifi.it}}}
\date{September 2025}

\begin{document}

\maketitle

\section*{Introduction}

\begin{figure}
    \centering
    \includegraphics[scale=0.2, angle=-90]{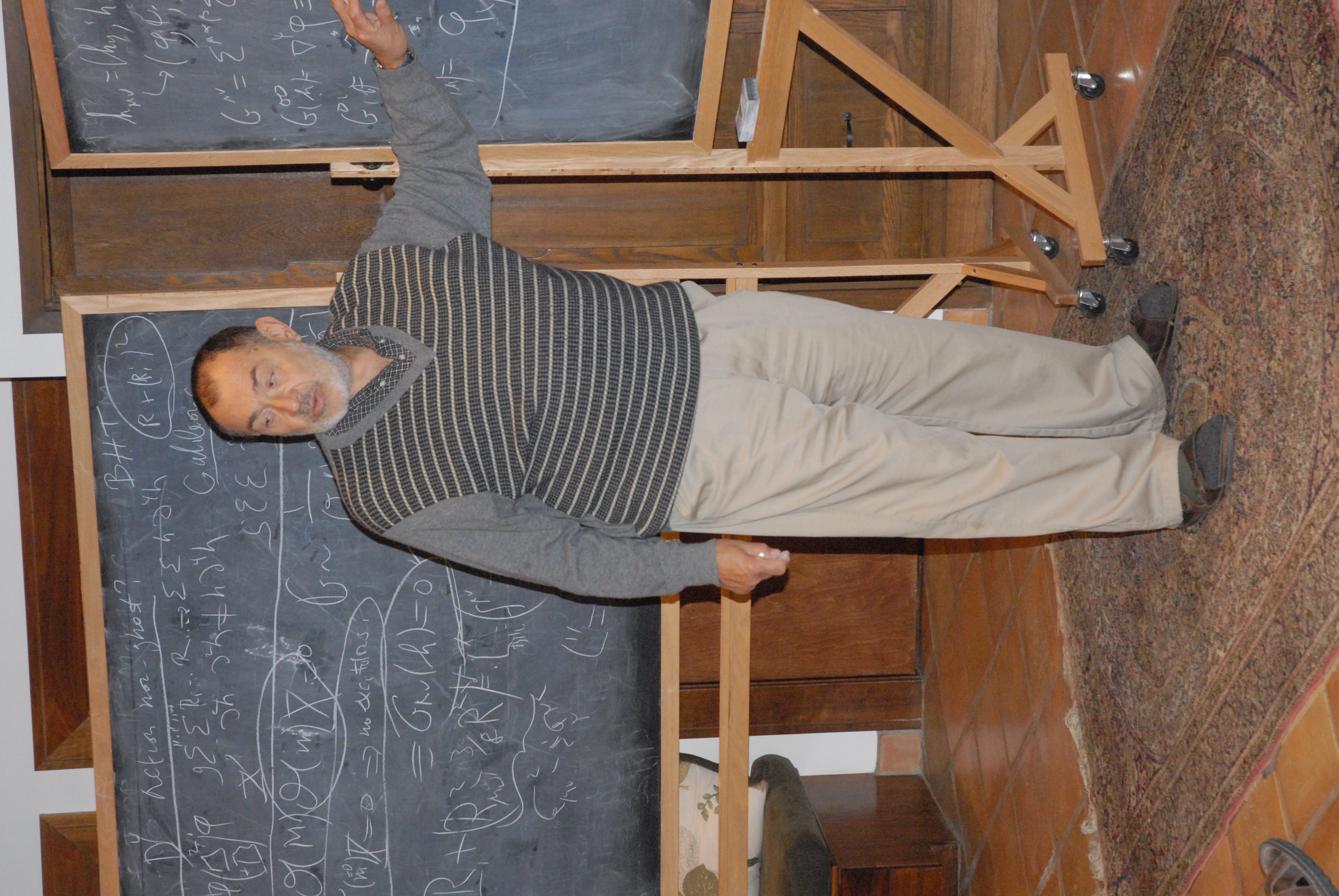}
    \caption{Stanley Deser. Credit: Anna Zytkow}
    \label{fig:photo}
\end{figure}

Stanley Deser (figure \ref{fig:photo}) was a towering figure in theoretical physics, whose influential contributions included the development of a Hamiltonian formulation of gravity \cite{Arnowitt:1959ah, Arnowitt:1962hi}, co-discovery of supergravity \cite{Deser:1976eh}, and explorations of gauge and gravity theories in three spacetime dimensions \cite{Deser:1981wh, Deser:1982vy, Deser:1983tn}. A comprehensive memoir of his scientific achievements is in preparation \cite{duff}.

On a personal level, Stanley was warm, unpretentious, charming, quick-witted and a raconteur par excellence.
One can still savor some of his anecdotes in his fascinating autobiography \cite{Deserbook}.

Stanley's contagious enthusiasm for physics and his congenial personality attracted over 100 collaborators during his career, from young students to distinguished senior researchers. We guest editors count ourselves fortunate to have been among them. Indeed, working with Stanley was invariably an intense, immersive, but thoroughly delightful experience. He was generous with encouragement, and he supported and took great interest in his collaborators.

It is therefore little wonder that so many prominent researchers have contributed to this special issue, making it a fitting tribute to Stanley's memory. The themes of the contributions fall roughly into the following four general areas:

\subsection*{Fields}

The article by Bern et al \cite{Bern:2019prr} reviews the duality between color and kinematics in scattering amplitudes of gauge and gravity theories. A general argument for trace anomaly matching between the unbroken and broken phases of a conformal field theory is given in the work by Schwimmer and Theisen \cite{Schwimmer:2023nzk}. Quantum field theory in two-dimensional de Sitter space, in particular the role of the discrete series unitary irreducible representation in the Hilbert space, is explored by Anninos et al \cite{Anninos:2023lin}. Dunne \cite{Dunne:2025wbq} extracts non-perturbative information about the tilted cusp anomalous dimension in $\mathcal{N}=4$ supersymmetric Yang-Mills theory using resurgent extrapolation and continuation methods based solely on perturbative data.
Vasiliev dedicated to Deser a paper \cite{Vasiliev:2025hfh} on supersymmetric higher-spin gauge theories in any dimension, which unfortunately could not be included in the special issue.

\subsection*{Gravity}

Nilsson and Pope \cite{Nilsson:2023ctq} analyze the phenomenon of de-Higgsing in the compactification of eleven-dimensional supergravity on a squashed S${}^7$. Fuentealba and Henneaux \cite{Fuentealba:2023fwe} investigate the Bondi–Metzner–Sachs group in 6 spacetime dimensions. 't Hooft \cite{tHooft:2024ydr} proposes a novel firewall transformation for black holes. A canonical analysis of $f$(Riemann) gravity is performed by Altas and Tekin \cite{Altas:2024utf}. Alessio and Di Vecchia \cite{Alessio:2024wmz} refine the computation of the waveform of the gravitational waves emitted by the scattering of two black holes. Zanelli and collaborators \cite{Giribet:2024nwg} investigate BPS defects in three-dimensional supergravity in anti-de Sitter space. Lovelock gravity and its maximally symmetric vacuum solutions are explored by Lindström and collaborators \cite{Devecioglu:2024qiu}. Castro and collaborators \cite{Bourne:2024ded} show how to incorporate massive spinning fields into the Euclidean path integral of three-dimensional quantum gravity. Porrati and Zaffaroni \cite{Porrati:2024zvi} discover a Higgs phenomenon in anti-de Sitter space. Marolf \cite{Marolf:2024jze} investigates the ensemble of theories describing spacetime wormholes in gravitational path integrals. Duff \cite{Duff:2024ikg} conjectures that an electric/magnetic duality is responsible for the vanishing of the beta function in gauged $N>4$ supergravity. Singletons in supersymmetric field theories and in supergravity are explored by
Samtleben and Sezgin \cite{Samtleben:2024zoy}. G\"unaydin \cite{Gunaydin:2024ihp} investigates massless conformal fields in ten dimensions, the minimal unitary representation of the conformal group $E_{7(-25)}$, and its connection to exceptional supergravity.
Howe and Lindström \cite{Howe:2024ojq} explore a loop-space formulation of an exotic linearised theory of superconformal gravity in six dimensions. Strominger and collaborators \cite{Melton:2024pre} give an explicit construction of the vacuum state for Klein space, which appears in the context of scattering amplitudes and flat space holography. Raclariu and collaborators \cite{He:2024vlp} study the infrared on-shell action of Einstein gravity in asymptotically flat spacetimes, and show that it is equal to the soft supertranslation charge, the shockwave effective action, and the soft effective action.
Waldron and collaborators \cite{Gover:2025yrh} introduce boundary curvature scalars on conformally compact manifolds. Woodard and Yesilyurt \cite{Woodard:2024zds} consider, after some reminiscences of Deser, the problem of implementing the Hamiltonian constraint for gravity + QED. 
Benetti Genolini and Murthy \cite{BenettiGenolini:2025jwe} show that the
Konstevich–Segal–Witten criterion is necessary but not sufficient for the allowability of complex
metrics contributing in the gravitational path integral to the superconformal index. Veneziano \cite{Veneziano:2025} discusses connections between
the ADM and BMS formalisms. Jacobson and Pulakkat \cite{Jacobson:2025zfu} analyze minimal Horava gravity  from the Hamiltonian point of view. Liu \cite{Liu:2025} considers breathing mode reductions of $D$-dimensional gravity theory to two dimensions and their relation to Jackiw-Teitelboim gravity. Giddings \cite{Giddings:2025xym} explores observables in quantum gravity and their mathematical structure.

\subsection*{Strings}

Schwarz \cite{Schwarz:2024une} briefly traces the 50-year history of string theory. Seibold and Tseytlin \cite{Seibold:2023zkz} compute the S-matrix of 2d massless fields starting from a membrane, and find that it is not integrable. Mourad and Sagnotti \cite{Mourad:2023loc} note the appearance of an effective orientifold from the breaking of supersymmetry in type IIB superstring theories. Herderschee and Maldacena \cite{Herderschee:2023pza} compute the three graviton amplitude in the matrix model for M-theory, and find agreement with the expected result form 
eleven-dimensional M-theory even at finite values of $N$. Bergshoeff et al \cite{Bergshoeff:2023rkk} investigate $p$-brane Galilean and Carrollian geometries and gravities. Witten \cite{Witten:2024yod} argues that a sigma model whose target space 
is the moduli space of instantons on ${\rm S}^3 \times {\rm S}^1$
has a large $\mathcal{N}=4$ superconformal symmetry, and may therefore be dual to type IIB superstring theory on ${\rm AdS}_3 \times {\rm S}^3 \times {\rm S}^3\times {\rm S}^1$. Hull \cite{Hull:2025yww} finds a consistent coupling of self-dual $p$-form gauge fields to self-dual branes.

\subsection*{Beyond}

Wilczek \cite{Wilczek:2023nmw} considers examples (taken from perception theory, rigid body mechanics, and quantum measurement) where ambiguity stemming from limited knowledge can be captured using negative values for quantities that are inherently positive. Raveh and Nepomechie \cite{Raveh:2023iyy} introduce $q$-analog qudit Dicke states, and investigate their properties.

\section*{Acknowledgments} 

We are grateful to the many friends and colleagues of Stanley who contributed to this special issue. 
The work of MH is partially supported by  FNRS-Belgium (convention IISN 4.4503.15) and by research funds from the Solvay Family. RN is supported in part by the National Science Foundation under grant PHY 2310594, and by a Cooper fellowship. DS is partially supported by the INFN grant 'GAST; Non perturbative effects in gauge and string theories.’

% \bibliographystyle{utphys}
% \bibliography{deser.bib}

\providecommand{\href}[2]{#2}\begingroup\raggedright\endgroup

\end{document}